\documentclass{article}
\usepackage{amsfonts}
\usepackage{makeidx}
\usepackage{latexsym,amsmath,amssymb,amscd}
\usepackage{makecell}
\usepackage{xcolor}
\usepackage{multirow}
\usepackage[all]{xy}
\usepackage{float}
\usepackage{longtable}

\allowdisplaybreaks
\newtheorem{theorem}{Theorem}

\begin{document}

\title{Integrable time-dependent central potentials}

\author{Antonios Mitsopoulos$^{1,a)}$ and Michael Tsamparlis$^{1,b)}$ \\
%EndAName
{\ \ }\\
$^{1}${\textit{Faculty of Physics, Department of
Astronomy-Astrophysics-Mechanics,}}\\
{\ \textit{University of Athens, Panepistemiopolis, Athens 157 83, Greece}}
\vspace{12pt} %EndAName
\\
$^{a)}$Author to whom correspondence should be addressed: antmits@phys.uoa.gr %EndAName
\\
$^{b)}$Email: mtsampa@phys.uoa.gr }
\date{}
\maketitle

\begin{abstract}
The integrable time-dependent central potentials that admit linear and quadratic first integrals other than those constructed from the angular momentum are determined. It is shown explicitly that previous answers to this problem are incomplete. The results are applied in order to find the integrable time-dependent oscillators, the integrable time-dependent generalized Kepler potentials, a class of integrable binary systems with variable mass, and the integrable Yukawa and interatomic potentials with time-dependent parameters. Finally, a new class of integrable potentials is integrated and the corresponding wavefunction is determined.
\end{abstract}
\bigskip


\section{Introduction}

\label{sec.intro}

There are various methods for the determination of the first integrals (FIs) of the time-dependent conservative systems
\begin{equation}
\ddot{q}^{a}= -U^{,a}(t,q)  \label{NP.0}
\end{equation}
where $t$ is the time, $q^{a}$ are the generalized coordinates with $a=1,2,...,n$, $n$ are the degrees of freedom of the system, $\dot{q}^{a}\equiv \frac{dq^{a}}{dt}$, $U(t,q)$ denotes the potential of the system, a comma indicates partial derivative, and the kinetic metric $\gamma_{ab}$ (i.e. the metric defined by the kinetic energy of the dynamical system) is used for lowering and raising indices. These methods include Noether's theorem, the Lie theory of extended groups, Ermakov's method, theory of canonical transformations, inverse Noether theorem, and the direct method (see e.g. \cite{Ermakov, Katzin 1973, Da Silva 1974, Djukic 1975, Fokas 1979, Katzin 1977, Katzin 1981, Katzin 1982, Leach 1981A, Leach 1985, Karlovini 2000, Horwood 2007, Tsamparlis 2020A, Tsamparlis 2020B}).

Mostly in the literature have been considered quadratic FIs (QFIs) of the form
\begin{equation}
I=K_{ab}(t,q)\dot{q}^{a}\dot{q}^{b}+K_{a}(t,q)\dot{q}^{a}+K(t,q)
\label{NP.1}
\end{equation}
where the Einstein summation convention is used and $K_{ab}(t,q), K_{a}(t,q), K(t,q)$ are symmetric tensor quantities. The case of linear FIs (LFIs) also included for $K_{ab}=0$. In the present work, we comment on two methods which have been used in order to determine the QFIs (\ref{NP.1}): a. The method of the Inverse Noether Theorem; and b. The direct method.

Concerning method a., one assumes that the dynamical equations (\ref{NP.0}) possess a regular Lagrangian $L(t,q,\dot{q})$ and uses the Inverse Noether theorem \cite{Djukic 1975, Tsamparlis 2020B} to associate to each QFI of the form (\ref{NP.1}) the generalized gauged Noether symmetry
\begin{equation}
\xi =0,\enskip\eta _{a}=-2K_{ab}\dot{q}^{b}-K_{a},\enskip f=-K_{ab}\dot{q}^{a}\dot{q}^{b}+K \label{NP.2}
\end{equation}
where $\mathbf{X}= \xi(t,q,\dot{q}) \frac{\partial}{\partial t} + \eta^{a}(t,q, \dot{q}) \frac{\partial}{\partial q^{a}}$ is the vector field that generates the Noether symmetry and $f$ denotes the Noether (or gauge) function. From (\ref{NP.2}), it follows that the generator $\eta^{a}$ is a linear function of the velocities; therefore, it has the general form%
\begin{equation}
\eta^{a} =A_{b}^{a}(t,q)\dot{q}^{b} +B^{a}(t,q)  \label{NP.4}
\end{equation}
where $A^{a}_{b}(t,q)$, $B^{a}(t,q)$ are tensor quantities. Subsequently, one substitutes $\eta^{a}$, $L$ in the generalized Killing equations \cite{VujanovicB} (see appendix) and finds
a system of partial differential equations (PDEs) whose solution determines the time-dependent potentials which admit QFIs.

Concerning method b., one requires the condition $dI/dt=0$ and, using the dynamical equations (\ref{NP.0}) to replace the term $\ddot{q}^{a}$ whenever it appears, finds a (different) set of PDEs which involves the unknown quantities $K_{ab}(t,q), K_{a}(t,q), K(t,q)$ and the time-dependent potential $U(t,q)$. Subsequently, from the dynamical equations (\ref{NP.0}), one reads the kinetic metric and, assuming that this metric is regular, uses standard results of Differential geometry, and `solves' the system of PDEs in terms of the geometric symmetries (collineations) of that metric.

In the present work, we address the problem of finding all time-dependent central potentials of regular conservative Newtonian systems that admit QFIs of the form (\ref{NP.1}); therefore, all such potentials that are (Liouville) integrable. Previous work on this problem has been done in \cite{Leach 1985} where the method of the inverse Noether theorem was used. It was concluded that there are three classes of time-dependent central potentials which were classified as Cases I, II, III given by the following expressions
\begin{equation}
V_{I}(t,r)= \frac{1}{2}\lambda (t)r^{2}  \label{NP.7}
\end{equation}%
\begin{equation}
V_{II}(t,r)=-\frac{\ddot{\phi}}{2\phi }r^{2}-\frac{\mu _{0}}{\phi }r^{-1}
\label{NP.8}
\end{equation}%
\begin{equation}
V_{III}(t,r)= -\frac{\ddot{\phi}}{2\phi }r^{2}+\phi ^{-2}F\left( \frac{r}{\phi }\right)  \label{NP.9}
\end{equation}
where $\phi(t)$ is an arbitrary smooth function, the frequency $\lambda(t)$ of the time-dependent oscillator may be either the function $\lambda_{1}(t)= \frac{K}{\phi^{4}} -\frac{\ddot{\phi}}{\phi}$ or the function $\lambda_{2}(t)= -\frac{\ddot{\phi}}{\phi}$, $\mu_{0},K$ are arbitrary constants, and $F$ is an arbitrary smooth function of its argument. For each potential the corresponding QFI is provided. We note that in fact this is one Case, because Cases I and II are subcases of Case III for $F=0$, $F=\frac{K}{2}\phi^{-2}r^{2}$ and $F=-\mu_{0} \frac{\phi}{r}$. As it will be shown, this result is a partial answer to the problem. It is important to note that in \cite{Leach 1985} the dynamical system is considered to be two-dimensional with variables $r, \theta$.

In a different approach \cite{LewLea 1982}, the authors considered one-dimensional (1d) Hamiltonian systems of the form
\begin{equation*}
H=\frac{1}{2}p^{2} +U(t,q) \label{NP.10}
\end{equation*}%
and, using the direct method for FIs of the form (\ref{NP.1}), determined all the time-dependent potentials $U(t,q)$ which admit QFIs. Obviously, in this case, central motion makes no sense. They concluded that these potentials are
\begin{equation}
U(t,r)= -\frac{\ddot{\rho}}{2\rho }q^{2}+\left( \alpha \frac{\ddot{\rho}}{\rho } -\ddot{\alpha}\right) q+\frac{1}{\rho ^{2}}G\left( \frac{q-\alpha }{\rho }\right)   \label{NP.3}
\end{equation}%
where $\rho (t), \alpha (t), G$ are arbitrary smooth functions of their arguments. The associated QFI is
\begin{equation}
I=\frac{1}{2}\left[ \rho (\dot{\rho}-\dot{\alpha}) -\dot{\rho}(q-\alpha )\right] ^{2} +G\left( \frac{q-\alpha}{\rho }\right). \label{NP.11}
\end{equation}
Equivalently, the potential (\ref{NP.3}) may be written as
\begin{equation}
U(t,r)=\frac{1}{2}\Omega ^{2}(t)q^{2}-F_{1}(t)q+\frac{1}{\rho ^{2}}\widetilde{G}\left( \frac{q-\alpha }{\rho }\right) \label{NP.12}
\end{equation}
where $\widetilde{G}$ is an arbitrary smooth function of its argument and the functions $\Omega(t), F_{1}(t), \rho(t), \alpha(t)$ satisfy the conditions
\begin{eqnarray}
\ddot{\rho} +\Omega^{2}(t)\rho -\frac{k}{\rho^{3}} &=& 0 \label{NP.13.1} \\
\ddot{\alpha} +\Omega^{2}(t)\alpha &=& F_{1}(t) \label{NP.13.2}
\end{eqnarray}
where $k$ is an arbitrary constant. In this notation, the associated QFI (\ref{NP.11}) becomes
\begin{equation}
I= \frac{1}{2} \left[ \rho (\dot{\rho} -\dot{\alpha}) -\dot{\rho} (q-\alpha) \right]^{2} +\frac{k}{2}\left( \frac{q-\alpha}{\rho} \right)^{2} +\widetilde{G}\left( \frac{q-\alpha}{\rho} \right). \label{NP.14}
\end{equation}

In the present work, we consider the problem of central motion and determine the time-dependent LFIs/QFIs using the direct method of \cite{LewLea 1982}. In order to do this, we have to reduce the degrees of freedom to one. We do this by means of the LFI of angular momentum.

The structure of the paper is as follows. In section \ref{sec.cenpot}, we reduce the degrees of freedom to the radial coordinate $r$, and, using the direct method, we compute the integrable time-dependent central potentials that admit LFIs (subsection \ref{sec.fx1}) and QFIs (subsection \ref{sec.fx2}). We collect our results in Theorem \ref{theorem}. In section \ref{applications}, we consider various applications of Theorem \ref{theorem}. Finally, in section \ref{conclusions}, we draw our conclusions.

\section{The integrable time-dependent central potentials $V(t,r)$}

\label{sec.cenpot}

The characteristic property of Newtonian central motion is that the angular momentum is conserved and the motion takes place on a plane normal to the angular momentum. On that plane we assume polar coordinates $(r,\theta)$ so that the Lagrangian becomes
\begin{equation}
L= \frac{1}{2} \left( \dot{r}^{2} +r^{2}\dot{\theta}^{2} \right) -V(t,r). \label{eq.cp3}
\end{equation}

The Euler-Lagrange (E-L) equations are
\begin{eqnarray}
\ddot{r} &=& r\dot{\theta}^{2} -\frac{\partial V}{\partial r} \label{eq.cp4a} \\
L_{3}&=& r^{2}\dot{\theta} \label{eq.cp4b}
\end{eqnarray}
where $L_{3}= r^{2}\dot{\theta}$ is the LFI of the angular momentum. Replacing $\dot{\theta}$ in (\ref{eq.cp4a}), we find the time-dependent second order ordinary differential equation (ODE) in the variable $r(t)$
\begin{equation}
\ddot{r}= \frac{L_{3}^{2}}{r^{3}} -\frac{\partial V(t,r)}{\partial r}. \label{eq.cp6}
\end{equation}
Solving equation (\ref{eq.cp6}), one finds a solution for $r(t)$ which when replaced in (\ref{eq.cp4b}) by integration gives the $\theta(t)$.

In what follows, we study the integrability of the time-dependent system
\begin{equation}
\ddot{r}= -\frac{\partial U}{\partial r} \label{eq.fx2}
\end{equation}
where the time-dependent potential
\begin{equation}
U(t,r)= \frac{L_{3}^{2}}{2r^{2}} +V(t,r). \label{eq.fx3}
\end{equation}
so that
\begin{equation}
V(t,r)= U(t,r) - \frac{L_{3}^{2}}{2r^{2}}. \label{eq.fx4}
\end{equation}
The problem of finding all the integrable potentials $U(t,r)$ has been solved in sections II and III of \cite{LewLea 1982} using the direct method. In the following, we update the approach of \cite{LewLea 1982}. Using (\ref{NP.1}) and the dynamical equations (\ref{NP.0}) to replace the term $\ddot{q}^{a}$ whenever it appears, condition $\frac{dI}{dt}=0$ leads to the following system of PDEs \cite{Tsamparlis 2020A}
\begin{eqnarray}
K_{(ab,c)} &=&0  \label{eq.cp8a} \\
K_{ab,t} +K_{(a,b)} &=&0  \label{eq.cp8b} \\
K_{a,t} +K_{,a} -2K_{ab}U^{,b} &=&0  \label{eq.cp8c} \\
K_{,t} -K_{a}U^{,a} &=&0  \label{eq.cp8d} \\
K_{a,tt} -2\left(K_{ab}U^{,b}\right)_{,t} +\left( K_{b}U^{,b} \right)_{,a} &=& 0 \label{eq.cp8e} \\
2\left( K_{[a|c|}U^{,c} \right)_{,b]} -K_{[a,b],t} &=& 0. \label{eq.cp8f}
\end{eqnarray}
Round/square brackets indicate symmetrization/antisymmetrization of the enclosed indices, and indices enclosed between vertical lines are overlooked by symmetrization or antisymmetrization symbols.

We note that the PDEs (\ref{eq.cp8a}), (\ref{eq.cp8b}) are purely geometric because they do not involve the potential $U(t,q)$. The PDE (\ref{eq.cp8a}) implies that $K_{ab}$ is a Killing tensor (KT) of order 2 (possibly zero) of $\gamma_{ab}$. Moreover, the last two equations (\ref{eq.cp8e}), (\ref{eq.cp8f}) express the integrability conditions for the scalar $K$.

Because the considered dynamical system (\ref{eq.fx2}) is 1d, the variable $q^{1}=r$ and the KT $K_{ab}$ is of the form $K_{11}=g_{1}(t)$. Therefore, the system of equations (\ref{eq.cp8a}) - (\ref{eq.cp8f}) becomes
\begin{eqnarray}
K_{11}&=& g_{1}(t) \label{eq.cp9.0} \\
K_{1}(t,r)&=& -\dot{g}_{1}r +g_{2}(t) \label{eq.cp9.1} \\
\frac{\partial K}{\partial r}&=& 2g_{1}\frac{\partial U}{\partial r} +\ddot{g}_{1}r -\dot{g}_{2} \label{eq.cp9.2} \\
\frac{\partial K}{\partial t}&=& \left(g_{2} -\dot{g}_{1} r \right) \frac{\partial U}{\partial r} \label{eq.cp9.3} \\
0&=& \left(\dot{g}_{1} r - g_{2}\right) \frac{\partial^{2} U}{\partial r^{2}} +2g_{1}\frac{\partial^{2}U}{\partial t\partial r} +3\dot{g}_{1}\frac{\partial U}{\partial r} +\dddot{g}_{1}r -\ddot{g}_{2} \label{eq.cp9.4}
\end{eqnarray}
where $g_{1}(t), g_{2}(t)$ are arbitrary smooth functions. The integrability condition (\ref{eq.cp9.4}) is ignored, because it is satisfied identically due to the PDEs (\ref{eq.cp9.2}), (\ref{eq.cp9.3}). Replacing $K_{1}, K_{11}$ given by (\ref{eq.cp9.0}), (\ref{eq.cp9.1}) in the expression (\ref{NP.1}) of the associated QFI we find
\begin{equation}
I= g_{1}\dot{r}^{2} +\left( g_{2} -\dot{g}_{1}r \right)\dot{r} +K(t,r). \label{eq.cp9.5}
\end{equation}

There remains the scalar $K$ and the corresponding time-dependent potential $U$ which shall be determined from the PDEs (\ref{eq.cp9.2}), (\ref{eq.cp9.3}). There are two cases to consider: a) $g_{1}(t)=0$ which provides the LFIs; and b) $g_{1}(t)\neq0$ which provides the QFIs.

\subsection{Case $g_{1}(t)=0$ (LFIs)}

\label{sec.fx1}

In this case, $K_{11}=0$, $K_{1}=g_{2}(t)\neq0$ and the PDEs (\ref{eq.cp9.2}), (\ref{eq.cp9.3}) become
\begin{eqnarray}
\frac{\partial K}{\partial r} &=& -\dot{g}_{2} \label{eq.fx5.1} \\
\frac{\partial K}{\partial t}&=& g_{2} \frac{\partial U}{\partial r}. \label{eq.fx5.2}
\end{eqnarray}
Integrating the PDE (\ref{eq.fx5.1}), we find that
\[
K(t,r)=-\dot{g}_{2}r +g(t)
\]
where $g(t)$ is an arbitrary function. Replacing this function in the PDE (\ref{eq.fx5.2}), we find the potential (see eq. (2.9) in \cite{LewLea 1982})
\begin{equation}
U(t,r)= -\frac{\ddot{g}_{2}}{2g_{2}}r^{2} +\frac{\dot{g}}{g_{2}}r. \label{eq.fx6.0}
\end{equation}
Substituting $U(t,r)$ in (\ref{eq.fx4}), we find the integrable time-dependent central potential
\begin{equation}
V(t,r)= -\frac{\ddot{g}_{2}}{2g_{2}}r^{2} +\frac{\dot{g}}{g_{2}}r -\frac{L_{3}^{2}}{2r^{2}}. \label{eq.fx6}
\end{equation}
This potential does not belong to any of the three classes found in \cite{Leach 1985} due to the additional term $\frac{\dot{g}}{g_{2}}r$. It is the sum of \newline
a. A repulsive time-dependent oscillator (term $-\frac{\ddot{g}_{2}}{2g_{2}}r^{2})$.\newline
b. An attractive Newton-Cotes potential (term $-\frac{L_{3}^{2}}{2r^{2}})$. \newline
c. A pure time-dependent central force (term $\frac{\dot{g}}{g_{2}}r$).

The associated LFI is
\begin{equation}
I= g_{2}\dot{r} -\dot{g}_{2}r +g \label{eq.fx7}
\end{equation}
which coincides with the LFI (2.11) of \cite{LewLea 1982}.

\subsection{Case $g_{1}(t)\neq0$ (QFIs)}

\label{sec.fx2}

In this case, equation (\ref{eq.cp9.2}) can be integrated and gives the potential (see eq. (3.8) in \cite{LewLea 1982})
\begin{equation}
U= \frac{K}{2g_{1}} -\frac{\ddot{g}_{1}}{4g_{1}}r^{2} +\frac{\dot{g}_{2}}{2g_{1}}r +g(t) \label{eq.fx8}
\end{equation}
where $g(t)$ is an arbitrary function. Replacing this potential in the PDE (\ref{eq.cp9.3}), we find the function (see eq. (3.14) in \cite{LewLea 1982})
\[
K= F\left( g_{1}^{-1/2}r +\frac{1}{2} \int g_{1}^{-3/2}g_{2}dt \right) +\frac{1}{4g_{1}} \left( \dot{g}_{1}r -g_{2} \right)^{2}
\]
where $F$ is an arbitrary smooth function of its argument.

Substituting the function $K$ in equation (\ref{eq.fx8}), we find the potential\footnote{We choose the function $g= -\frac{g_{2}^{2}}{8 g_{1}^{2}}$ so as not to have an additive function of $t$ in the potential.}
\begin{equation}
U= \left[ \frac{1}{8} \left(\frac{\dot{g}_{1}}{g_{1}} \right)^{2} -\frac{\ddot{g}_{1}}{4g_{1}} \right] r^{2} +\frac{1}{2g_{1}} \left( \dot{g}_{2} -g_{2} \frac{\dot{g}_{1}}{2g_{1}} \right)r +\frac{1}{2g_{1}} F\left( g_{1}^{-1/2}r +\frac{1}{2} \int g_{1}^{-3/2}g_{2}dt \right).
\label{eq.fx8.1}
\end{equation}
Replacing $U$ in equation (\ref{eq.fx4}), we obtain the integrable time-dependent central potential
\begin{equation}
V(t,r)= \left[ \frac{1}{8} \left(\frac{\dot{g}_{1}}{g_{1}} \right)^{2} -\frac{\ddot{g}_{1}}{4g_{1}} \right] r^{2} +\frac{1}{2g_{1}} \left( \dot{g}_{2} -g_{2} \frac{\dot{g}_{1}}{2g_{1}} \right)r +\frac{1}{2g_{1}} F\left( g_{1}^{-1/2}r +\frac{1}{2} \int g_{1}^{-3/2}g_{2}dt \right) -\frac{L_{3}^{2}}{2r^{2}}. \label{eq.fx9}
\end{equation}

The associated QFI (\ref{eq.cp9.5}) is
\begin{equation}
I= g_{1}\dot{r}^{2} +(g_{2} -\dot{g}_{1}r)\dot{r} +F\left( g_{1}^{-1/2}r +\frac{1}{2} \int g_{1}^{-3/2}g_{2}dt \right) +\frac{1}{4g_{1}} \left( \dot{g}_{1}r -g_{2} \right)^{2}. \label{eq.fx10}
\end{equation}

We note that for
\begin{equation}
g_{1}=\frac{\phi^{2}}{2}, \enskip g_{2}=0, \enskip F= \bar{F} +\frac{L_{3}^{2}\phi^{2}}{2r^{2}} \label{eq.fx10.0}
\end{equation}
where $F, \bar{F}$ are functions of the same argument, we recover the Case III potentials of \cite{Leach 1985} as a special case of the potential (\ref{eq.fx9}).

Using the inverse Noether theorem, we have that for the QFI (\ref{eq.fx10}) the associated gauged Noether symmetry (\ref{NP.2}) is
\begin{eqnarray}
\eta_{1}&=& -2g_{1}\dot{r} +\dot{g}_{1}r -g_{2} \label{eq.fx10.1} \\
f&=& -g_{1}\dot{r}^{2} +F\left( g_{1}^{-1/2}r +\frac{1}{2} \int g_{1}^{-3/2}g_{2}dt \right) +\frac{1}{4g_{1}} \left( \dot{g}_{1}r -g_{2} \right)^{2}. \label{eq.fx10.2}
\end{eqnarray}
These results coincide with the ones of \cite{Leach 1985} provided one replaces $\dot{\theta}$ from $L_{3}= r^{2}\dot{\theta}$.

We collect the results in the following Theorem.

\begin{theorem} \label{theorem}
The integrable time-dependent central Newtonian potentials $V(t,r)$ with angular momentum $L_{3}$ are the following: \newline
a. The potentials $V(t,r)= -\frac{\ddot{g}_{2}}{2g_{2}}r^{2} +\frac{\dot{g}}{g_{2}}r -\frac{L_{3}^{2}}{2r^{2}}$ which admit the LFIs $I=g_{2}\dot{r}-\dot{g}_{2}r +g$ where $g_{2}(t)\neq 0, g(t)$ are arbitrary functions. \newline
b. The potentials $V(t,r)=\left[ \frac{1}{8}\left( \frac{\dot{g}_{1}}{g_{1}}%
\right) ^{2}-\frac{\ddot{g}_{1}}{4g_{1}}\right] r^{2}+\frac{1}{2g_{1}}\left(
\dot{g}_{2}-g_{2}\frac{\dot{g}_{1}}{2g_{1}}\right) r+\frac{1}{2g_{1}}F\left(
g_{1}^{-1/2}r+\frac{1}{2}\int g_{1}^{-3/2}g_{2}dt\right) -\frac{L_{3}^{2}}{%
2r^{2}}$ which admit the QFIs $I=g_{1}\dot{r}^{2}+(g_{2}-\dot{g}_{1}r)\dot{r}%
+F\left( g_{1}^{-1/2}r+\frac{1}{2}\int g_{1}^{-3/2}g_{2}dt\right) +\frac{1}{%
4g_{1}}\left( \dot{g}_{1}r-g_{2}\right)^{2}$ where $g_{1}(t)\neq0, g_{2}(t)$ are arbitrary functions.
\end{theorem}

It can be shown that our results are in agreement with those of \cite{LewLea 1982}.

\section{Applications of Theorem \protect\ref{theorem}}

\label{applications}

From Theorem \ref{theorem}, it follows that it is possible to classify all integrable time-dependent central Newtonian potentials in just two cases according to if they admit a. LFIs or b. QFIs. In this section, we consider various applications of Theorem \ref{theorem}.

\subsection{The time-dependent oscillator}

\label{sec.app1}

In this case, the potential is of the form $V=-\omega(t)r^{2}$ where $\omega(t)$ is an arbitrary function. The LFIs and the QFIs are as follows.

a. For $g(t)=0, L_{3}=0 \implies \theta(t)=const$, we have the time-dependent potential
\begin{equation}
V= -\frac{\ddot{g}_{2}}{2g_{2}}r^{2} \label{osc.1}
\end{equation}
with the LFI
\begin{equation}
I= g_{2}\dot{r} -\dot{g}_{2}r. \label{osc.2}
\end{equation}

b. For $g_{2}=0$ and $F= \frac{c_{0}}{2g_{1}}r^{2} +L_{3}^{2} \frac{g_{1}}{r^{2}}$, where $c_{0}$ is an arbitrary constant, we obtain the time-dependent potential
\begin{equation}
V= -\left[ \frac{\ddot{g}_{1}}{4g_{1}} -\frac{1}{8}\left( \frac{\dot{g}_{1}}{g_{1}} \right)^{2} -\frac{c_{0}}{4g_{1}^{2}} \right] r^{2} \label{osc.3}
\end{equation}
with the QFI
\begin{equation}
I= g_{1}\left( \dot{r}^{2} + \frac{L_{3}^{2}}{r^{2}} \right) -\dot{g}_{1}r\dot{r} +\frac{\dot{g}_{1}^{2}}{4g_{1}} r^{2} +\frac{c_{0}}{2g_{1}} r^{2}. \label{osc.4}
\end{equation}
If we replace $L_{3}=r^{2}\dot{\theta}$, the QFI (\ref{osc.4}) is the sum of the diagonal components of the Jauch-Hill-Fradkin tensor. If in addition $g_{1}=\frac{\phi^{2}}{2}$ and $c_{0}=\frac{K}{2}$, where $\phi(t)$ is an arbitrary function and $K$ a constant, we derive the Case I QFI of \cite{Leach 1985}
\begin{equation}
I= \frac{1}{2} \left( \phi \dot{r} -\dot{\phi}r \right)^{2} +\frac{1}{2}r^{2}\phi^{2}\dot{\theta}^{2} +\frac{K}{2\phi^{2}} r^{2}. \label{osc.5}
\end{equation}

The above results for the time-dependent oscillator coincide with the ones of Theorem 6.2 of \cite{Katzin 1977}.

\subsection{The time-dependent generalized Kepler potential}

\label{sec.app2}

In this case, $V= -\frac{\omega(t)}{r^{\nu}}$ where $\nu$ is an arbitrary non-zero constant.

a. No new integrable potentials.

b. For $g_{1}=\frac{\phi^{2}}{2}$, $g_{2}=0$ and $F= \bar{F} +\frac{L_{3}^{2}\phi^{2}}{2r^{2}}$, where $F, \bar{F}$ are functions of the same argument and $\phi(t)$ an arbitrary smooth function, we find the Case III potential of \cite{Leach 1985}
\begin{equation}
V= -\frac{\ddot{\phi}}{2\phi} r^{2} +\phi^{-2} \bar{F} \left( \frac{r}{\phi} \right) \label{gen.1}
\end{equation}
with the QFI
\begin{equation}
I= \frac{1}{2} (\phi\dot{r} -r\dot{\phi})^{2} +\frac{L_{3}^{2} \phi^{2}}{2r^{2}} +\bar{F}\left( \frac{r}{\phi} \right) = \frac{1}{2} (\phi\dot{r} -r\dot{\phi})^{2} +\frac{1}{2} \phi^{2} r^{2} \dot{\theta}^{2} +\bar{F}\left( \frac{r}{\phi} \right) \label{gen.2}
\end{equation}
where we replaced $L_{3}=r^{2}\dot{\theta}$.

We choose
\[
\bar{F}\left( \frac{r}{\phi} \right)= k_{1}\frac{r^{2}}{\phi^{2}} -\frac{k\phi^{\nu}}{r^{\nu}}
\]
with
\[
\phi= \sqrt{b_{0}+ b_{1}t +b_{2}t^{2}}, \enskip k_{1}= \frac{b_{0}b_{2}}{2} -\frac{b_{1}^{2}}{8}
\]
where $\nu, k, b_{0}, b_{1}, b_{2}$ are arbitrary constants. Then the potential (\ref{gen.1}) becomes the integrable time-dependent generalized Kepler potential
\begin{equation}
V= -\frac{\omega_{\nu}(t)}{r^{\nu}}, \enskip \omega_{\nu}= k\left(b_{0} + b_{1}t + b_{2}t^{2} \right)^{\frac{\nu-2}{2}} \label{gen.3}
\end{equation}
which admits the QFI
\begin{equation}
J_{\nu}=  (b_{0} + b_{1}t + b_{2}t^{2}) \left[ \frac{1}{2} \left( \dot{r}^{2} +r^{2}\dot{\theta}^{2} \right) - \frac{\omega_{\nu}}{r^{\nu}} \right] -\frac{b_{1} + 2b_{2}t}{2} r\dot{r} +\frac{b_{2} r^{2}}{2}. \label{gen.4}
\end{equation}
This result is in accordance with Propositions 1 and 2 of \cite{Mathematics 2021}.

We note that in the case of the standard Kepler potential (i.e. $\nu=1$) for $k_{1}=0 \implies b_{1}^{2} -4b_{0}b_{2}=0$, the `frequency' $\omega_{1}= \frac{k}{\sqrt{b_{0}+ b_{1}t+b_{2}t^{2}}}$ degenerates into the well-known result $\omega_{1}= \frac{k}{a_{0} +a_{1}t}$ where $a_{0}, a_{1}$ are constants (see Theorem 2.1 of \cite{Katzin 1982} and Proposition 3 of \cite{Mathematics 2021}).

\subsection{Integrating the equations of motion for the case a. potentials of Theorem \ref{theorem}}

\label{sec.app4}

In Theorem \ref{theorem} we found the new class of time-dependent integrable central potentials (\ref{eq.fx6}) which admit the LFIs (\ref{eq.fx7}).

Using the transformation $R= \frac{g_{2}}{r}$, we compute
\[
\dot{R}= \frac{\dot{g}_{2}r -g_{2}\dot{r}}{r^{2}}.
\]
Replacing this in the LFI (\ref{eq.fx7}), we find the solution
\begin{equation}
r(t)= g_{2} \int \frac{I -g}{g_{2}^{2}} dt +c \label{eq.cl3}
\end{equation}
where $c$ is an integration constant.

Substituting (\ref{eq.cl3}) in $L_{3}= r^{2}\dot{\theta}$, we find that
\begin{equation}
\theta(t)= \int \frac{L_{3}}{r^{2}(t)}dt +\theta_{0} \label{eq.cl4}
\end{equation}
where $\theta_{0}$ is another integration constant.

\subsection{Integrating the Scr\"{o}dinger equation for an integrated central potential}

\label{sec.app3}

We consider a special class of integrable central potentials with fixed angular momentum $L_{3}$ generated from case b. of Theorem \ref{theorem} for
\[
g_{1}= \frac{\phi^{2}}{2}, \enskip g_{2}=0, \enskip F= -k\frac{\phi}{r}
\]
where $\phi(t)$ is an arbitrary non-zero function and $k\neq0$ is an arbitrary constant. Replacing in the defining formula, we obtain the potential
\begin{equation}
V(t,r)= -\frac{\ddot{\phi}}{2\phi}r^{2} -\frac{k}{\phi}r^{-1} -\frac{L_{3}^{2}}{2r^{2}}. \label{eq.e1}
\end{equation}
The associated QFI is
\begin{equation}
I= \frac{1}{2} \left( \phi \dot{r} -\dot{\phi}r \right)^{2} -k\frac{\phi}{r}. \label{eq.e2}
\end{equation}

By introducing the transformation $R= \frac{\phi}{r}$ we compute
\[
\dot{R}= \frac{\dot{\phi}r -\phi \dot{r}}{r^{2}}
\]
and replacing in the QFI (\ref{eq.e2}) we find that
\begin{equation}
\frac{dR}{R^{2}\sqrt{2(I+kR)}}= \pm \frac{dt}{\phi^{2}}. \label{eq.e3}
\end{equation}
Using the LFI of angular momentum, we find the following orbit
\[
L_{3}= r^{2} \dot{\theta} \implies d\theta = L_{3} R^{2} \frac{dt}{\phi^{2}} \implies \int d\theta= \pm L_{3}\int \frac{dR}{\sqrt{2(I+kR)}} \implies
\]
\begin{equation}
\theta= \pm \frac{L_{3}}{k} \sqrt{2 \left( I +\frac{k\phi}{r} \right)} +\theta_{0} \label{eq.e4}
\end{equation}
where $\theta_{0}$ is an integration constant.

In \cite{Leach 1985} it has been shown that the wavefunction for the Case III potential (\ref{NP.9}) is given by
\begin{equation}
\psi(r,\theta,t)= |\phi|^{-1/2} e^{\frac{i}{2\hbar}\frac{\dot{\phi}}{\phi}r^{2}} \bar{\psi} \left( \phi^{-1}r, \theta, T(t) \right) \label{eq.schr12}
\end{equation}
where\footnote{We note that in eq. (6.14) of \cite{Leach 1985} the second term into the brackets should be multiplied by 2.}
\begin{equation}
\bar{\psi}(R,\Theta,T)= e^{-i\lambda T/\hbar} e^{im\Theta} R^{-1/2} A(R) \label{eq.schr10}
\end{equation}
$\lambda, m$ are arbitrary constants, $R= \phi^{-1}r$, $\Theta= \theta$, $T(t)= \int \phi^{-2}dt \implies \dot{T}= \phi^{-2}$ and $A(R)$ is an arbitrary smooth function which satisfies the second order ODE (see eq. (6.6) of \cite{Leach 1985})
\begin{equation}
\frac{d^{2}A}{dR^{2}} +\left( \frac{2\lambda}{\hbar^{2}} -\frac{2}{\hbar^{2}}F -\frac{m^{2} -\frac{1}{4}}{R^{2}} \right) A =0. \label{eq.schr11}
\end{equation}
The parameter $\hbar$ is the Planck constant.

We observe that for
\[
F\left( \frac{r}{\phi} \right)= -k\phi r^{-1} -\frac{L_{3}^{2} \phi^{2}}{2r^{2}} \implies F(R)= -\frac{k}{R} -\frac{L_{3}^{2}}{2R^{2}}
\]
the potential (\ref{NP.9}) reduces to the potential (\ref{eq.e1}) for which the orbit equation has been found. For this choice of $F$ the ODE (\ref{eq.schr11}) becomes
\begin{equation}
\frac{d^{2}A}{dR^{2}} +\left( \frac{2\lambda}{\hbar^{2}} +\frac{2k}{\hbar^{2}R} -\frac{m^{2} -\frac{1}{4} -\frac{L_{3}^{2}}{\hbar^{2}}}{R^{2}} \right) A =0. \label{eq.schr13}
\end{equation}
From Table 22.6, p. 781 of \cite{Abramowitz} we find that (\ref{eq.schr13}) admits the solution
\begin{equation}
A(R)= e^{-R/2} R^{\frac{a+1}{2}} L^{(a)}_{b}(R) \label{eq.schr14}
\end{equation}
where $L^{(a)}_{b}(R)$ is a generalized Laguerre polynomial and the constants $\lambda, k, m$ are given by
\begin{equation}
\lambda=-\frac{\hbar^{2}}{8}, \enskip k= \frac{\hbar^{2}}{4} \left( 2b +a +1 \right), \enskip m^{2}= \frac{a^{2}}{4} +\frac{L_{3}^{2}}{\hbar^{2}}. \label{eq.schr15}
\end{equation}
Substituting the above results in (\ref{eq.schr12}) we find the wavefunction
\begin{equation}
\psi(r,\theta,t)= |\phi|^{-1/2} e^{\frac{i}{2\hbar}\frac{\dot{\phi}}{\phi} r^{2}} e^{i\hbar T(t)/8} e^{i \sqrt{\frac{a^{2}}{4} +\frac{L_{3}^{2}}{\hbar^{2}}} \theta} \phi^{1/2} r^{-1/2} e^{-\frac{r}{2\phi}} \left( \frac{r}{\phi} \right)^{\frac{a+1}{2}} L^{(a)}_{b}(\phi^{-1}r) \label{eq.schr16}
\end{equation}
provided the defining constant of the potential (\ref{eq.e1}) is $k= \frac{\hbar^{2}}{4} \left( 2b +a +1 \right)$.

\subsection{The integrable two-body problem with variable mass}

\label{sec.app5}

In Newtonian Physics, the motion of a point of variable mass $m(t)$ is described by the dynamical equation \cite{Hadjidemetriou 1963, Plastino 1992}
\begin{equation}
m(t)\ddot{\mathbf{R}}= \mathbf{F} + \sum_{i} \dot{m}_{i} \mathbf{u}_{i} \label{eq.bin1}
\end{equation}
where $\mathbf{R}$ is the position vector of $m(t)$ with respect to (wrt) a fixed frame of reference, $\mathbf{F}$ denotes the external forces, $\mathbf{u}_{i}$ is the relative velocity of the escaping mass from the $i$th-point of the surface of $m(t)$, and $\dot{m}_{i}$ is the rate of loss of mass from the $i$th-point. If the loss is continuous, the summation symbol may be replaced by a double integral over the whole surface around $m$.

It is said that \textbf{the loss of mass is isotropic} iff $\sum_{i} \dot{m}_{i} \mathbf{u}_{i}=0$. For example, this is the case when the stars lose mass by radiation.

The two-body problem with variable mass \cite{Hadjidemetriou 1963} (e.g. a binary system of stars) consists of two point masses $m_{1}(t), m_{2}(t)$ with only gravitational attraction. Assuming isotropic loss of mass, the equations of motion wrt a fixed inertial frame are
\begin{equation}
m_{1}(t) \ddot{\mathbf{r}}_{1}= \frac{G m_{1}m_{2}}{r^{2}} \hat{\mathbf{r}}, \enskip m_{2}(t) \ddot{\mathbf{r}}_{2}= -\frac{G m_{1}m_{2}}{r^{2}} \hat{\mathbf{r}} \label{eq.bin2}
\end{equation}
where $G$ is the gravitational constant, $\mathbf{r}= \mathbf{r}_{2} -\mathbf{r}_{1}$ is the relative position of the star with mass $m_{2}$ wrt the other star with mass $m_{1}$, $r= \|\mathbf{r}\|$ and the unit vector $\hat{\mathbf{r}}= \frac{\mathbf{r}}{r}$. From equations (\ref{eq.bin2}) we find the dynamical equation
\begin{equation}
\ddot{\mathbf{r}}= -\frac{Gm(t)}{r^{2}} \hat{\mathbf{r}} \label{eq.bin3}
\end{equation}
where $m(t)\equiv m_{1}(t) +m_{2}(t)$ is the total mass of the binary system. If $m(t)=const$, the system is called closed and the orbit is a well-known conic section. The potential driving the system is the time-dependent Kepler potential
\begin{equation}
V(t,r)= -\frac{\omega(t)}{r} \label{eq.bin4}
\end{equation}
where the `frequency' $\omega(t)= Gm(t)$. The problem which has been around for a long time was the determination of the mass $m(t)$ so that the potential (\ref{eq.bin4}) is integrable. It has been found  \cite{Gylden, Mestschersky 1893, Mestschersky 1902, Prieto 1997, Rahoma 2009} that this potential is integrable for the following $m(t)$:
\[
m_{I}(t)= \frac{1}{a_{0}+a_{1}t}, \enskip m_{II}(t)= \frac{1}{\sqrt{b_{0}+b_{1}t}}, \enskip m_{III}(t)= \frac{1}{\sqrt{b_{0}+b_{1}t+b_{2}t^{2}}}.
\]
Using Theorem \ref{theorem}, we derive these functions easily. They are subcases of the `frequency' $\omega_{\nu}$ of the integrable time-dependent generalized Kepler potential (\ref{gen.3}) for $\nu=1$ and $k=G$. Indeed, we have
\[
\omega_{1}(t)= Gm(t) \implies \frac{G}{\sqrt{b_{0} +b_{1}t +b_{2}t^{2}}} =Gm(t) \implies m(t)= \frac{1}{\sqrt{b_{0}+b_{1}t+b_{2}t^{2}}}= m_{III}(t).
\]
For $b_{2}=0$ we obtain the mass $m_{II}(t)$ and for vanishing discriminant $b_{1}^{2} -4b_{0}b_{2}=0$ the mass $m_{I}(t)$.

\subsection{Time-dependent integrable Yukawa and interatomic potentials}

\label{sec.app6}

In plasma physics, solid-state physics and nuclear physics, the following types of central potentials are widely used:
\bigskip

1. The Yukawa type potentials
\begin{equation}
V(r)= A\frac{e^{-Br}}{r} \label{eq.Y1}
\end{equation}
where $A, B$ are arbitrary constants. This type of potentials describes the screened Coulomb potential \cite{Debye 1923} generated around a positive charged particle into a neutral fluid (e.g. a plasma of electrons in a background of heavy positive charged ions \cite{Bellan}), and also models successfully the neutron-proton interaction \cite{Yukawa 1935}.
\bigskip

2. The interatomic pair potentials \cite{Jones 1924A, Jones 1924B}
\begin{equation}
V(r)= \frac{A}{r^{m}} -\frac{B}{r^{n}} \label{eq.A1}
\end{equation}
where $A, B, m, n$ are arbitrary positive constants. These central potentials manifest between the atoms of diatomic molecules. From (\ref{eq.A1}) we observe that they consist of a repulsive term $\frac{A}{r^{m}}$ and an attractive term $-\frac{B}{r^{n}}$. The most well-known potential of this form is the Lennard-Jones potential \cite{LJ potential} in which $m=12$ and $n=6$.
\bigskip

Using Theorem \ref{theorem}, we shall answer to the following problem: \newline
\emph{Assume the parameters $A, B$ of the potentials (\ref{eq.Y1}), (\ref{eq.A1}) to be time-dependent. Then find for which functions $A(t), B(t)$ the resulting potentials are integrable.}

The case b. potentials of Theorem \ref{theorem} for
\[
g_{2}=0, \enskip F= -\frac{c_{1}}{4g_{1}}r^{2} +\frac{L_{3}^{2}g_{1}}{r^{2}} +\bar{F} \left( g_{1}^{-1/2} r \right)
\]
where $c_{1}$ is an arbitrary constant and $\bar{F}$ a smooth function of its argument, reduce to the integrable potentials
\begin{equation}
V(t,r)= -\left[ \frac{\ddot{g}_{1}}{4g_{1}} -\frac{1}{8}\left( \frac{\dot{g}_{1}}{g_{1}} \right)^{2} +\frac{c_{1}}{8g_{1}^{2}} \right] r^{2} +\frac{1}{2g_{1}} \bar{F} \left( g_{1}^{-1/2} r \right) \label{eq.R1}
\end{equation}
which admit the QFIs
\begin{equation}
I= \left(\dot{r}^{2} +\frac{L_{3}^{2}}{r^{2}} \right) g_{1} -\dot{g}_{1}r\dot{r} -\frac{c_{1}}{4g_{1}} r^{2} +\frac{\dot{g}_{1}^{2}}{4g_{1}} r^{2} +\bar{F} \left( g_{1}^{-1/2} r \right). \label{eq.R2}
\end{equation}

From (\ref{eq.R1}) we see that integrable time-dependent central potentials of the form
\begin{equation}
V(t,r)= \frac{1}{2g_{1}} \bar{F} \left( g_{1}^{-1/2} r \right) \label{eq.R3}
\end{equation}
exist only when
\begin{equation}
g_{1}(t)=b_{0}+b_{1}t+b_{2}t^{2}, \enskip c_{1}=b_{1}^{2} -4b_{2}b_{0} \label{eq.R4}
\end{equation}
where $b_{0}, b_{1}, b_{2}$ are arbitrary constants. For special choices of the function $\bar{F}$ we obtain the required integrable potentials:
\bigskip

1. $\bar{F}= \frac{2k\exp{\left( -g_{1}^{-1/2}r \right)}}{g_{1}^{-1/2}r}$ where $k$ is an arbitrary constant.

We find the new integrable time-dependent Yukawa type potentials
\begin{equation}
V(t,r)=\frac{k}{\sqrt{b_{0}+b_{1}t+b_{2}t^{2}}} \frac{e^{-\frac{r}{\sqrt{b_{0}+b_{1}t+b_{2}t^{2}}}}}{r} \label{eq.R5}
\end{equation}
where $A(t)= \frac{k}{\sqrt{b_{0}+b_{1}t+b_{2}t^{2}}}$ and $B(t)= \frac{1}{\sqrt{b_{0}+b_{1}t+b_{2}t^{2}}}$.
\bigskip

2. $\bar{F}= \frac{2k_{1}g_{1}^{m/2}}{r^{m}} -\frac{2k_{2}g_{1}^{n/2}}{r^{n}}$ where $k_{1}, k_{2}$ are arbitrary constants.

We find the new integrable time-dependent interatomic pair potentials
\begin{equation}
V(t,r)= \frac{k_{1}(b_{0}+b_{1}t+b_{2}t^{2})^{\frac{m-2}{2}}}{r^{m}} -\frac{k_{2}(b_{0}+b_{1}t+b_{2}t^{2})^{\frac{n-2}{2}}}{r^{n}} \label{eq.R6}
\end{equation}
where $A(t)= k_{1}(b_{0}+b_{1}t+b_{2}t^{2})^{\frac{m-2}{2}}$ and $B(t)= k_{2}(b_{0}+b_{1}t+b_{2}t^{2})^{\frac{n-2}{2}}$.

\section{Conclusions}

\label{conclusions}

Using the LFI of angular momentum and the direct method, we have managed to compute the integrable time-dependent central potentials. These potentials are widely used in all branches of Physics. It is remarkable that so divergent potentials can be squeezed into two simple classes, i.e. the ones which admit LFIs and the ones which admit QFIs. One may ask: Why in \cite{Leach 1985} not all the integrable time-dependent central Newtonian potentials were found? The reason is that in \cite{Leach 1985} the generalized Killing equations, instead of the dynamical equations, were used. Indeed, the former result from the Noether condition which is written in the form $A(t,q,\dot{q}) +B_{a}(t,q,\dot{q})\ddot{q}^{a}=0$ by requiring $B_{a}=0$ and
do not use the dynamical equations to replace the $\ddot{q}^{a}$.

\section*{Appendix: The generalized Killing equations}

\label{con.mot.subsec.killing.2}

Consider a Lagrangian dynamical system that admits a generalized Noether symmetry generated by $\xi = \xi(t, q, \dot{q})$, $\eta^i = \eta^i(t, q, \dot{q})$ with Noether function $f = f(t, q, \dot{q})$. Then the Noether condition
\begin{equation}
\mathbf{X}^{[1]}(L) +\dot{\xi}L =\dot{f} \label{app.1}
\end{equation}
where $\mathbf{X}^{[1]}= \xi \partial_{t} +\eta^{i} \partial_{q^{i}} +(\dot{\eta}^{i} -\dot{q}^{i}\dot{\xi}) \partial_{\dot{q}^{i}}$, becomes
\begin{align*}
0 &= \xi \frac{\partial L}{\partial t} + \eta^i \frac{\partial L}{\partial q^i} + \left( \frac{\partial \eta^i}{\partial t} + \dot{q}^j \frac{\partial \eta^i}{\partial q^j} - \dot{q}^i \frac{\partial \xi}{\partial t} - \dot{q}^i \dot{q}^j \frac{\partial \xi}{\partial q^j} \right) \frac{\partial L}{\partial \dot{q}^i} + L \left( \frac{\partial \xi}{\partial t} + \dot{q}^i \frac{\partial \xi}{\partial q^i} \right) - \frac{\partial f}{\partial t} - \dot{q}^i \frac{\partial f}{\partial q^i} + \\
& \quad + \ddot{q}^j \left[ \frac{\partial \xi}{\partial \dot{q}^j} L + \left( \frac{\partial \eta^i}{\partial \dot{q}^j} - \dot{q}^i \frac{\partial \xi}{\partial \dot{q}^j} \right) \frac{\partial L}{\partial \dot{q}^i} - \frac{\partial f}{\partial \dot{q}^j} \right].
\end{align*}
Since this condition must be satisfied identically for all $\ddot{q}^{i}$, it follows
\begin{eqnarray}
\xi \frac{\partial L}{\partial t} + \eta^i \frac{\partial L}{\partial q^i} + \left( \frac{\partial \eta^i}{\partial t} + \dot{q}^j \frac{\partial \eta^i}{\partial q^j} - \dot{q}^i \frac{\partial \xi}{\partial t} - \dot{q}^i \dot{q}^j \frac{\partial \xi}{\partial q^j} \right) \frac{\partial L}{\partial \dot{q}^i} + L \left( \frac{\partial \xi}{\partial t} + \dot{q}^i \frac{\partial \xi}{\partial q^i} \right) &=& \frac{\partial f}{\partial t} + \dot{q}^i \frac{\partial f}{\partial q^i} \label{con.mot.eq.kil3} \\
\frac{\partial \xi}{\partial \dot{q}^j} L + \left( \frac{\partial \eta^i}{\partial \dot{q}^j} - \dot{q}^i \frac{\partial \xi}{\partial \dot{q}^j} \right) \frac{\partial L}{\partial \dot{q}^i} &=& \frac{\partial f}{\partial \dot{q}^j}. \label{con.mot.eq.kil4}
\end{eqnarray}
These equations are \textbf{the generalized Killing equations}. They coincide with conditions (A1), (A2) in the appendix of \cite{Leach 1985}. We note that we have $n+2$ unknowns $\xi$, $\eta^{i}$, $f$, but only $n+1$ equations. Therefore, there
is a free degree of freedom which is removed by a gauge condition (usually the condition $\xi =0$).

\end{document}